# Fuzzy and interval finite element method for heat conduction problem

Sarangam Majumdar[1], Sukanta Nayak[2] and S. Chakraverty[3]

Department of Mathematics, National Institute of Technology, Rourkela, Odisha -769008, India

**Abstract**

Traditional finite element method is a well-established method to solve various problems of science and engineering. Different authors have used various methods to solve governing differential equation of heat conduction problem. In this study, heat conduction in a circular rod has been considered which is made up of two different materials viz. aluminum and copper. In earlier studies parameters in the differential equation have been taken as fixed (crisp) numbers which actually may not. Those parameters are found in general by some measurements or experiments. So the material properties are actually uncertain and may be considered to vary in an interval or as fuzzy and in that case complex interval arithmetic or fuzzy arithmetic has to be considered in the analysis. As such the problem is discretized into finite number of elements which depend on interval/fuzzy parameters. Representation of interval/fuzzy numbers may give the clear picture of uncertainty. Hence interval/fuzzy arithmetic is applied in the finite element method to solve a steady state heat conduction problem. Application of fuzzy finite element method in the said problem gives fuzzy system of linear equations in general. Here new methods have also been proposed to handle such type of fuzzy system of linear equations. Corresponding results are computed and has been reported here.

1. Corresponding Author.; E-mail : majumdarsarangam@yahoo.in (SarangamMajumdar)

2. E-mail: sukantgacr@gmail.com (SukantaNayak)

3. E-mail : sne_chak@yahoo.com (SnehashishChakraverty )

**Nomenclature**

$K_x$ :             Thermal Conduttivity

$T_1, T_2, T_3, T_4, T_5$ :    All of them are temperature at different nodes

$q_x$ :             Heat Flux

$Q$ :              Internal heat generation rate

$U$ :              Internal energy

$A$ :              Area of cross section.

$[K]$ :            Conductance matrix

$\{f_Q\}$ :         Nodal vector

$\{f_g\}$ :         Gradient boundary conditions at the element nodes

$\underline{x}$ :            Left value of the interval

$\bar{x}$ :             Right value of the interval

## 1. Introduction

Conduction of heat means transfer of heat energy within the body due to the temperature gradient. Heat spontaneously flows from a body having higher temperature to lower temperature. But in absence of external driving fluxes it approaches to thermal equilibrium. There are two types of conduction such as steady and unsteady state. Steady state conduction is a form of conduction where the temperature differences deriving by the conduction remains constant and it is independent of time. The steady state heat conduction problem is well known and its solution by exact method has been solved earlier [1]. The analysis may be difficult when heat transfer through a complicated domain. Various numerical techniques are proposed for these types of problems viz. finite difference method, finite volume method and finite element method [2, 3]. Magnus et al. [2] used finite difference method in his paper to model and solve the governing ground water flow rates flow direction and hydraulic heads through an aquifer. Muhieddine et al. [3] described one dimensional phase change problem. They have used vertex centered finite volume method to solve the problem. In view of the above literature, it reveals that the traditional finite element method may easily be used where the parameters or the values are exact that is in crisp form. But in actual practice the values may be in a region of possibility or we can say the values are uncertain. In general uncertainty may be found from limited knowledge where it is impossible to exactly describe the existing state, vagueness, no specificity and dissonance etc. or we can find it if the probabilistic description of sampling variables are not available. These uncertain parameters give uncertain model predictions. Now the uncertainty can be reduced by appropriate experiments still it also give the variability in the parameters. To handle such variability several probabilistic methods have been introduced. Monte Carlo method is used as an alternative method for this type of problem. Then finite element perturbation method is used by Nicolaï and De Baerdemaeker [4] and Nicolaï[ 5] for heat conduction problem with uncertain physical parameters. Further Nicolaï [6, 7] found the temperature in heat conduction problem for randomly varying parameters with respect to time. They have used a variance propagation technique to calculate the mean and covariance of the temperatures. As we cannot find always a probabilistic description for the uncertain parameters some effort has been taken to tackle this type of problem. Hence we need the help of interval/fuzzy analysis for handling these types of data. We have used the interval arithmetic [8, 9, 10] which is described in third section of this paper. Then we present the traditional finite element procedure [11, 12, 13] for solving the problem by taking these parameters as interval. But it is a tedious task to solve by this process. Again there is a chance of occurring weak solutions. Next the interval finite element technique is described for the said problem. Finite element method in the present problem turns in to a system of linear equations which is solved by a proposed iterative method. Further new methods have also been used to solve the interval system of linear equations. Next fuzzy parameters are handled using the alpha cut techniques and finally we apply the fuzzy finite element method [14, 15, 16] .The proposed techniques for system of interval and fuzzy linear system of equations are used to solve a steady state heat conduction problem [17]. Finally we have given the numerical results and compared the different methods.

## 2. Traditional Finite Element Formulation

The principle of conservation of energy viz.

$$E_{in} + E_{generated} = E_{increase} + E_{out} \tag{1}$$

satisfies the following heat transfer equation

$$q_x A dt + QA dx dt = \Delta U + \left(q_x + \frac{\partial q_x}{\partial x} dx\right) A dt \qquad (2)$$

where, $q_x$ is the heat flux across boundary, $Q$ is the internal heat generation rate, $U$ is the internal energy and $A$ is the area of cross section.

The one dimension steady state heat conduction equation [17] may be written as

$$K_x \frac{d^2 T}{dx^2} + Q = 0. \qquad (3)$$

Traditional finite element formulation [17] for equation (3) can easily be obtained as

$$[K]\{T\} = \{f_Q\} + \{f_g\} \qquad (4)$$

where, $[K]$ is the conductance matrix, $\{f_Q\}$ is the nodal vector, $\{f_g\}$ is the gradient boundary conditions at the element nodes.

The stiffness matrix or conductance matrix for one element for traditional finite element method is given by $\frac{k_x A}{l}\begin{bmatrix} 1 & -1 \\ -1 & 1 \end{bmatrix}$, where $k_x$ is the thermal conductivity, $A$ is the area of cross-section and $l$ is the length of the cylinder. As such if the domain is divided into $n$ elements, the form of stiffness matrix is $\begin{bmatrix} 1 & -1 & 0 & \cdots & 0 \\ -1 & 2 & -1 & \ddots & \vdots \\ 0 & -1 & 2 & \ddots & 0 \\ \vdots & \ddots & \ddots & \ddots & -1 \\ 0 & \cdots & 0 & -1 & 2 \end{bmatrix}$.

## 3. Interval Arithmetic

The interval form of the parameters may be written as

$$[\underline{x}, \bar{x}] = \{x : x \in \mathfrak{R}, \underline{x} \leq x \leq \bar{x}\}$$

where $\underline{x}$ is the left value and $\bar{x}$ is the right value of the interval respectively. We define $m = \dfrac{\underline{x} + \bar{x}}{2}$ is the centre and $w = \bar{x} - \underline{x}$ is the width of the interval $[\underline{x}, \bar{x}]$.

Let $[\underline{x}, \bar{x}]$ and $[\underline{y}, \bar{y}]$ be two elements then the following arithmetic are well known [8]

(i) $[\underline{x}, \bar{x}] + [\underline{y}, \bar{y}] = [\underline{x} + \underline{y}, \bar{x} + \bar{y}]$

(ii) $[\underline{x}, \bar{x}] - [\underline{y}, \bar{y}] = [\underline{x} - \bar{y}, \bar{x} - \underline{y}]$

(iii) $[\underline{x}, \bar{x}] \times [\underline{y}, \bar{y}] = [\min\{\underline{x}\underline{y}, \underline{x}\bar{y}, \bar{x}\underline{y}, \bar{x}\bar{y}\}, \max\{\underline{x}\underline{y}, \underline{x}\bar{y}, \bar{x}\underline{y}, \bar{x}\bar{y}\}]$

(iv) $[\underline{x}, \bar{x}] \div [\underline{y}, \bar{y}] = [\min\{\underline{x} \div \underline{y}, \underline{x} \div \bar{y}, \bar{x} \div \underline{y}, \bar{x} \div \bar{y}\}, \max\{\underline{x} \div \underline{y}, \underline{x} \div \bar{y}, \bar{x} \div \underline{y}, \bar{x} \div \bar{y}\}]$

## 4. Fuzzy number and alpha cut

Let $X$ denote a universal set. Then, the membership function $\mu_A$ by which a fuzzy set $A$ is usually defined as the form $\mu_A : X \to [0,1]$, where $[0,1]$ denotes the interval of real numbers from 0 to 1. Such a function is called a membership function and the set defined by it is called a fuzzy set. A fuzzy number is a convex, normalized fuzzy set $A \subseteq R$ which is piecewise continuous and has the functional value $\mu_A(x) = 1$, where $x \in X$ at precisely one element. Different types of fuzzy numbers are there. These are triangular fuzzy number, trapezoidal fuzzy number and Gaussian fuzzy number etc. Here we have discussed the said problem using triangular fuzzy number only. The membership function for triangular fuzzy number is as below.

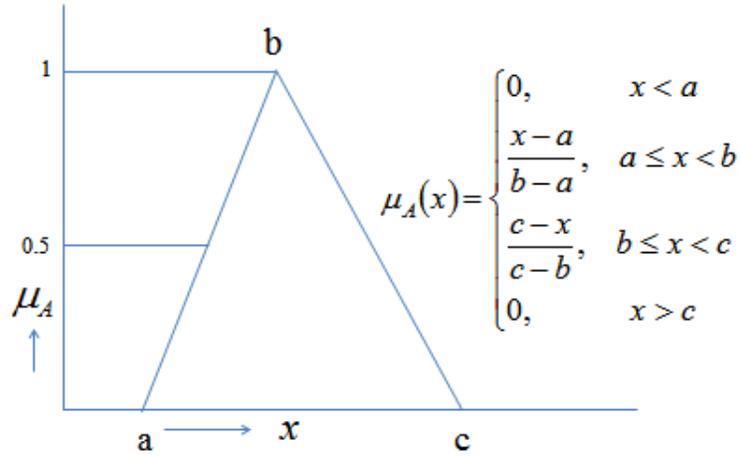

Fig.1.Triangular fuzzy number [a, b, c]

Alpha cut ($\alpha-$ cut) is an important concept of fuzzy set. Given $\alpha \in [0,1]$ then the alpha cut for above triangular fuzzy number $[a,b,c]$ may be written as $[a+(b-a)\alpha, c-(c-b)\alpha]$.

## 5. Interval and Fuzzy Finite Element Method

In this method the crisp values are replaced by interval/fuzzy and then proceeding like traditional finite element method we get a linear system of interval equations as given below.

$$\begin{bmatrix} [a_{11},b_{11}] & [a_{12},b_{12}] & \cdots & [a_{1n},b_{1n}] \\ [a_{21},b_{21}] & [a_{22},b_{22}] & \cdots & [a_{2n},b_{2n}] \\ \vdots & \vdots & \vdots & \vdots \\ [a_{n1},b_{n1}] & [a_{n2},b_{n2}] & \cdots & [a_{nn},b_{nn}] \end{bmatrix} \begin{Bmatrix} [x_1,y_1] \\ [x_2,y_2] \\ \vdots \\ [x_n,y_n] \end{Bmatrix} = \begin{Bmatrix} [c_1,d_1] \\ [c_2,d_2] \\ \vdots \\ [c_n,d_n] \end{Bmatrix}$$
(5)

where the members in an interval are both either positive or negative.

The system of equations in equation (5) can be solved by direct elimination method which we generally do for the exact data. Now interval values are operated through the interval arithmetic rules. Because of the more computation involved in this procedure it is a difficult task to perform and the margin of uncertainty increases drastically. Hence we have to choose methods which give better results. To overcome the above difficulties (to a certain extent) we now propose two new methods.

**Method-1**

Equation (5) can be solved by iterative scheme like Gauss-Seidel numerical method. The new form of representation of Equation (5) is as follows

$$[x_1, y_1]^{(k+1)} = \frac{[c_1, d_1] - [a_{12}, b_{12}][x_2, y_2]^{(k)} - [a_{13}, b_{13}][x_3, y_3]^{(k)} - \cdots - [a_{1n}, b_{1n}][x_n, y_n]^{(k)}}{[a_{11}, b_{11}]}$$

$$[x_2, y_2]^{(k+1)} = \frac{[c_2, d_2] - [a_{21}, b_{21}][x_1, y_1]^{(k)} - [a_{23}, b_{23}][x_3, y_3]^{(k)} - \cdots - [a_{2n}, b_{2n}][x_n, y_n]^{(k)}}{[a_{22}, b_{22}]}$$

...
...
...

$$[x_n, y_n]^{(k+1)} = \frac{[c_n, d_n] - [a_{n1}, b_{n1}][x_1, y_1]^{(k)} - [a_{n2}, b_{n2}][x_2, y_2]^{(k)} - \cdots - [a_{n(n-)1}, b_{n(n-)1}][x_{n-1}, y_{n-1}]^{(k)}}{[a_{nn}, b_{nn}]}$$

(6)

The above procedure will be more efficient if we replace the vector $[x, y]^{(k)}$ in the right side of Equation (6) element by element. So Equation (6) can be written as

$$[x_1, y_1]^{(k+1)} = \frac{[c_1, d_1] - [a_{12}, b_{12}][x_2, y_2]^{(k)} - [a_{13}, b_{13}][x_3, y_3]^{(k)} - \cdots - [a_{1n}, b_{1n}][x_n, y_n]^{(k)}}{[a_{11}, b_{11}]}$$

$$[x_2, y_2]^{(k+1)} = \frac{[c_2, d_2] - [a_{21}, b_{21}][x_1, y_1]^{(k+1)} - [a_{23}, b_{23}][x_3, y_3]^{(k)} - \cdots - [a_{2n}, b_{2n}][x_n, y_n]^{(k)}}{[a_{22}, b_{22}]}$$

...
...
...

$$[x_n, y_n]^{(k+1)} = \frac{[c_n, d_n] - [a_{n1}, b_{n1}][x_1, y_1]^{(k+1)} - [a_{n2}, b_{n2}][x_2, y_2]^{(k+1)} - \cdots - [a_{n(n-)1}, b_{n(n-)1}][x_{n-1}, y_{n-1}]^{(k)}}{[a_{nn}, b_{nn}]}$$

(7)

**Method-2**

Equation (5) can be written as

$$\sum_{j=1}^{n}(a_{ij}+\alpha_{ij})x_j = c_j + \beta_j, \quad i=1,2,3,...,n$$

where $\alpha_{ij} = \dfrac{b_{ij}-a_{ij}}{N}$ and $\beta_j = \dfrac{d_j-c_j}{N}, N \in [1,\infty)$.

(8)

Now using any crisp method we can solve Equation (8) and the solution for this system of equations may be written as below

$$\left[\min\left\{\lim_{n\to 1}x_1, \lim_{n\to\infty}x_1\right\}, \max\left\{\lim_{n\to 1}x_1, \lim_{n\to\infty}x_1\right\}\right], \left[\min\left\{\lim_{n\to 1}x_2, \lim_{n\to\infty}x_2\right\}, \max\left\{\lim_{n\to 1}x_2, \lim_{n\to\infty}x_2\right\}\right],$$
$$..., \left[\min\left\{\lim_{n\to 1}x_n, \lim_{n\to\infty}x_n\right\}, \max\left\{\lim_{n\to 1}x_n, \lim_{n\to\infty}x_n\right\}\right].$$

As mentioned earlier in case of fuzzy finite element methods we have to convert the fuzzy parameters into alpha cut set and then solve it by the proposed method as above. In this case we get a set of solutions for different values of alpha. Thus we get the solutions in term of triangular fuzzy number where the left values of the interval and the right values of the interval all together gives a pricewise continuous function.

The example problem has been solved by crisp finite element, interval finite element and fuzzy finite element method with the help of the above two proposed methods which are discussed in the subsequent sections.

**6. Problem and Numerical Results**

We have taken a circular rod (Fig.1) having an outside diameter of 60mm, length of 1m and perfectly insulated on its circumference. The left half of the cylinder is aluminum and the right half is copper having the material thermal conductivity $200 W/m-°C$ and $389 W/m-°C$ respectively. The extreme right end of the cylinder is maintained at a temperature of $80°C$, while the left end is subjected to a heat input rate $4000 W/m^2$ as given in [17].

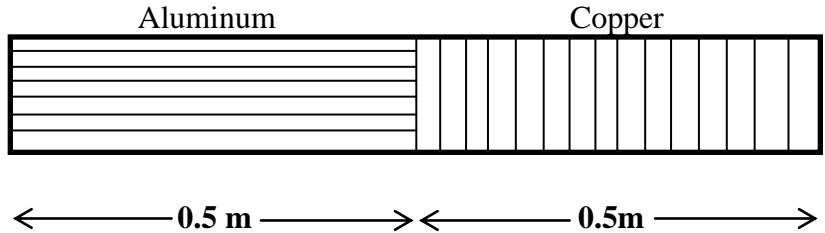

Fig.2.Model diagram for circular rod which is discretized into two equal parts having length 0.5meach.

Here the interval values for thermal conductivity viz. of aluminum has been taken as [199, 201] and for copper as [388,390]. The interval value for the heat input rate is taken as [3999, 4001]. The Problem is solved first by using the direct method and proposed iteration method (method-1) for four element discretization of circular rod. Corresponding results are given in table 1.

**Table 1.** Comparison of results using traditional and proposed iteration method.

| Temperatures (in degree Celsius) | Usual method | | | Method-1 | | |
|---|---|---|---|---|---|---|
| | Left | Right | Centre | left | right | Centre |
| $T_1$ | 87.11 | 104.54 | 95.825 | 89.01 | 101.40 | 95.205 |
| $T_2$ | 82.96 | 98.52 | 90.74 | 84.43 | 96.04 | 90.235 |
| $T_3$ | 80.47 | 90.54 | 85.505 | 81.14 | 89.28 | 85.21 |
| $T_4$ | 80.03 | 85.50 | 82.765 | 80.15 | 85.58 | 82.865 |
| $T_5$ | 80 | 80 | 80 | 80 | 80 | 80 |

Next two element and four element discretization have been considered. The proposed method (method 2) has been used to solve the final linear equations obtained from the finite element analysis. Corresponding results from Method 2 along with the results from traditional finite element method for crisp values are presented in Tables 2 to 5 respectively. Results of Table 2 and Table 3 are obtained for two and four element discretization respectively by taking the interval values for thermal conductivity only. Then both thermal conductivity and heat input rate are taken as interval values at a time for the analysis and the results with two and four element discretization respectively are given in Tables 4 and 5.

**Table 2.** Results from two element discretization with thermal conductivity in interval form

| Temperatures (in degree Celsius) | Traditional Finite element method (crisp) | Method-2, Interval finite element method | | |
|---|---|---|---|---|
| | | Left | Right | Centre |
| $T_1$ | 95.1414 | 95.0785 | 95.2049 | 95.1417 |
| $T_2$ | 85.1414 | 85.1282 | 85.1546 | 85.1414 |
| $T_3$ | 80 | 80 | 80 | 80 |

**Table 3.** Results from four element discretization with thermal conductivity in interval form

| Temperatures (in degree Celsius) | Traditional Finite element method (crisp) | Method-2, Interval finite element method | | |
|---|---|---|---|---|
| | | Left | Right | Centre |
| $T_1$ | 95.1414 | 95.0785 | 95.2049 | 95.1417 |
| $T_2$ | 90.1414 | 90.1033 | 90.1798 | 90.14155 |
| $T_3$ | 85.1414 | 85.1282 | 85.1546 | 85.1414 |
| $T_4$ | 82.5707 | 82.5641 | 82.5773 | 82.5707 |
| $T_5$ | 80 | 80 | 80 | 80 |

**Table 4.** Results from two element discretization with thermal conductivity and heat input rate both in interval form

| Temperatures (in degree Celsius) | Traditional Finite element method (crisp) | Method-2, Interval finite element method | | |
|---|---|---|---|---|
| | | Left | Right | Centre |
| $T_1$ | 95.1414 | 95.0822 | 95.2011 | 95.14165 |
| $T_2$ | 85.1414 | 85.1295 | 85.1534 | 85.14145 |
| $T_3$ | 80 | 80 | 80 | 80 |

**Table 5.** Results from four element discretization with thermal conductivity and heat input rate both in interval form

| Temperatures (in degree Celsius) | Traditional Finite element method (crisp) | Method-2, Interval finite element method | | |
|---|---|---|---|---|
| | | Left | Right | Centre |
| $T_1$ | 95.1414 | 95.0822 | 95.2011 | 95.14165 |
| $T_2$ | 90.1414 | 90.1059 | 90.1772 | 90.14155 |
| $T_3$ | 85.1414 | 85.1295 | 85.1534 | 85.14145 |
| $T_4$ | 82.5707 | 82.5647 | 82.5767 | 82.5707 |
| $T_5$ | 80 | 80 | 80 | 80 |

Now the values of thermal conductivity for aluminum and copper are taken in interval form with large width viz. [197.5, 202.5] and [386.5, 391.5] respectively and the heat input rate as [3997.5, 4002.5]. Obtained results with four element discretization are incorporated in Table 6.

**Table 6.** Result from four element discretization with large width for thermal conductivity of aluminum and copper heat input rate

| Temperatures (in degree Celsius) | Traditional Finite element method (crisp) | Method-2, Interval finite element method | | |
|---|---|---|---|---|
| | | Left | Right | Centre |
| $T_1$ | 95.1414 | 94.9945 | 95.2917 | 95.1431 |
| $T_2$ | 90.1414 | 90.0531 | 90.2315 | 90.1423 |
| $T_3$ | 85.1414 | 85.1117 | 85.1714 | 85.14155 |
| $T_4$ | 82.5707 | 82.5559 | 82.5857 | 82.5708 |
| $T_5$ | 80 | 80 | 80 | 80 |

Next let us consider the value of thermal conductivity for aluminum as TFN [199, 200, 201], the value of thermal conductivity for copper as TFN [388, 389, 390] and the heat input rate as TFN [3999, 4000, 4001]. Obtained results in terms of fuzzy plot are depicted in Figs. 3 to 6.

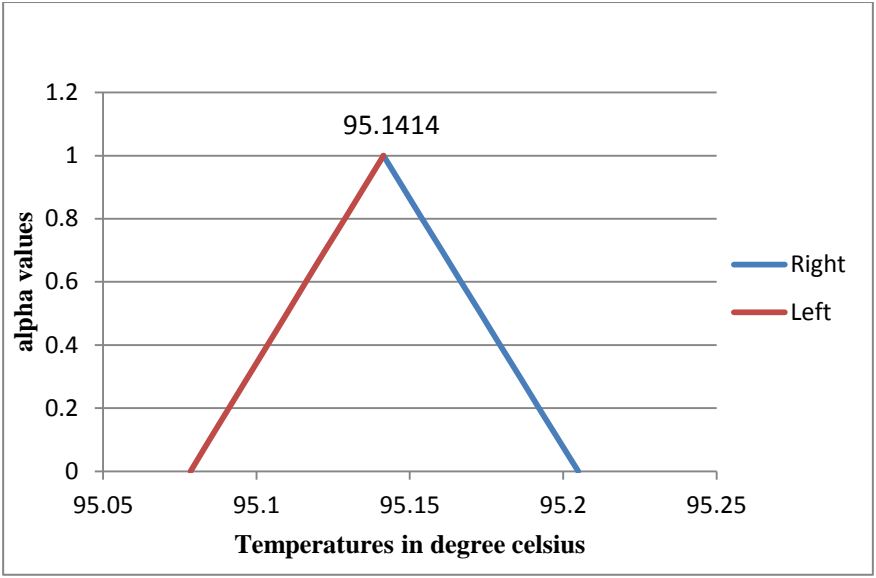

Fig. 3. Fuzzy plot for Temperature $T_1$

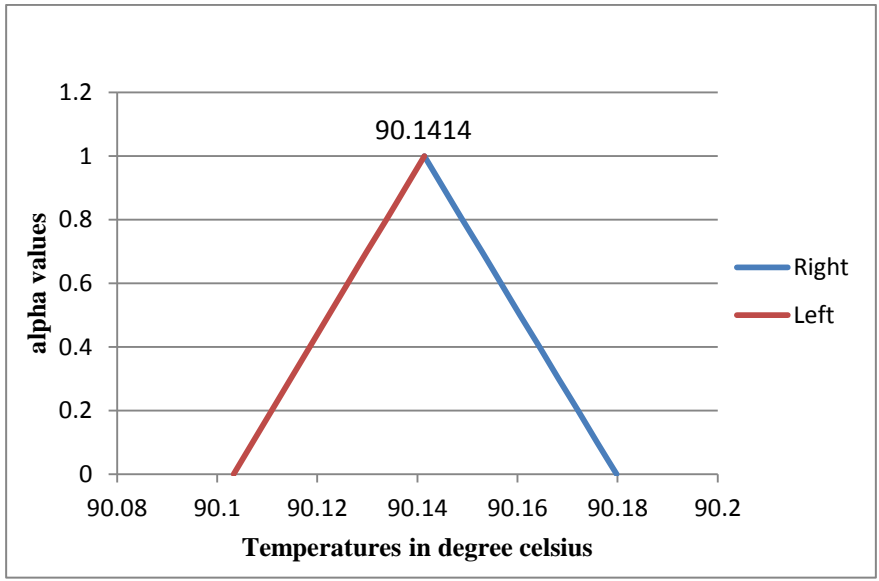

Fig.4. Fuzzy plot for Temperature $T_2$

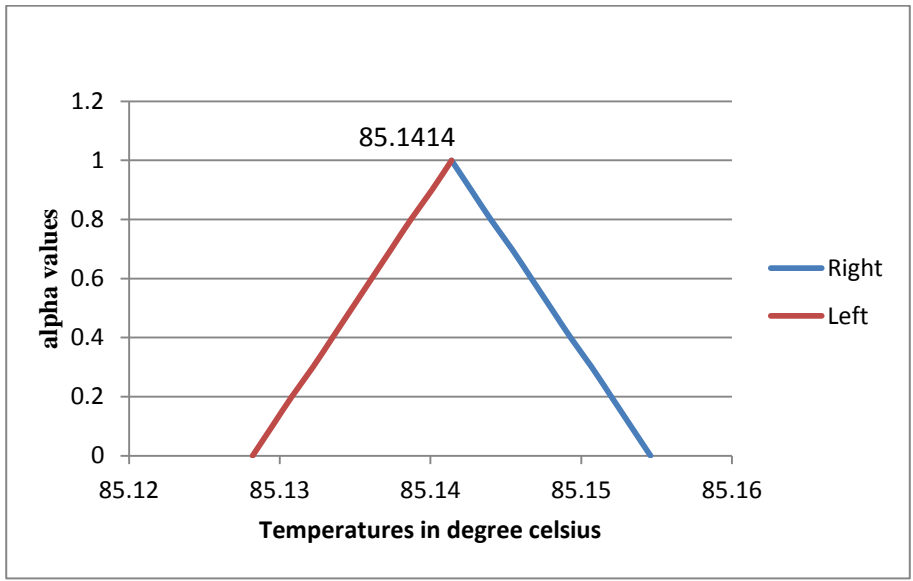

Fig. 5. Fuzzy plot for Temperature $T_3$

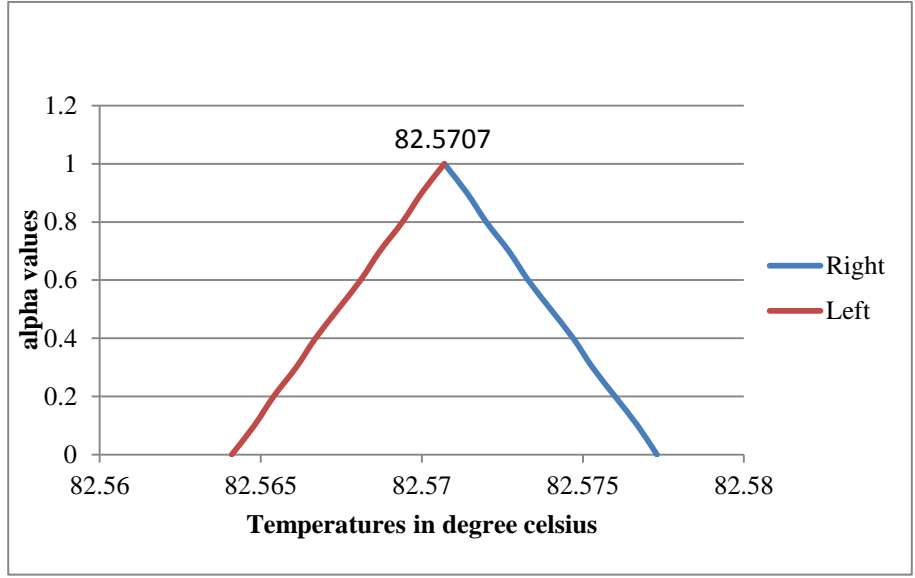

Fig. 6. Fuzzy plot for Temperature $T_4$

## 7. Discussion

Here we found that the marginal error caused due to the uncertain parameter used in the said problem becomes more for the usual interval finite element method which is shown in Table1. Whereas using method 1 we get a comparatively better result. The possibility of the solution set is decreased, which gives a better picture to predict the temperatures at the nodal point of the domain (Fig.2). Again if we consider the centre value of the solution set and comparing the two results of Table 1, method 1 gives a better approximation compare to the results of traditional finite element method (in crisp form). This may be seen by looking into Table 1 and Table 5 (column2).

Then the results of Table 2 and Table 3 for two element and four element discretization respectively give a comparative study about method 2 and the traditional finite element method (in crisp). Method 2 also gives a better approximation to the exact result. Again from Table 4 and Table 5 we are getting the idea of temperature distribution of domain more clearly in comparison with the usual method and method 1. Thus the possibility of temperature distribution along the rod for method 2 is far better than method 1. The centered result for method 2 (in Table 5) gives a sharp approximation to the exact result than the method 1 (in Table 1). We may also note from Table 6 that if the width of the interval is more for the uncertain parameter, method 2 gives a clear idea of temperature and it is more suitable to visualize the behavior of temperature compare to other methods.

Considering the uncertain parameters as fuzzy we have presented the result of the said problem in Figs. 3 to 6. When the value of alpha becomes zero the fuzzy results change to the interval form and for the value of alpha as one, the result changes into crisp form. In Figs. 3 to 6 we get a series of narrow and peak distribution of temperatures which reflect better solution for the said problem and are very close to the solution obtained from the traditional finite element method with crisp parameters.

## 8. Conclusion

This paper investigates in detail one-dimension steady state heat conduction problem with uncertain parameters. The parameters involved in the governing equation dictate the solution result. It is well known that the involved parameters cannot be obtained in general exactly or in crisp form. So the same are considered as uncertain in term of fuzzy/interval. As such the investigation is done by using fuzzy/interval finite element method to solve the test problem. Two proposed methods are introduced to find the numerical solution of the said problem with uncertainty. Corresponding results are given and compared with the known result in the special cases. Although the example problem seems to be simple but the main aim of this study is to develop fuzzy/interval finite element method (F/I FEM). It is worth mentioning that the F/I FEM for the said type of problems converts the problem into fuzzy/interval linear system of equations. Accordingly new methods are proposed which may very well be applicable to other complicated problems also.

**Acknowledgements**

The authors would like to thank BRNS (Board of Research in Nuclear Sciences)**,** Department of Atomic Energy, (DAE), Govt. of India for providing fund to do this work.